\documentclass[a4paper,10pt]{article}

\usepackage[utf8]{inputenc}
\usepackage{fullpage}
\usepackage{amsmath}
\usepackage{bm}
\usepackage{amsfonts}
\usepackage{amssymb}
\usepackage[pdftex]{hyperref}
\usepackage{graphicx}
\usepackage{algorithmic}
\usepackage{algorithm}
\usepackage{subfigure}
\usepackage[numbers,sort&compress]{natbib}
\usepackage{authblk}

\title{Discrimination and characterization of {P}arkinsonian rest tremors by analyzing long-term correlations and multifractal signatures}

\author[1]{Lorenzo Livi\thanks{llivi@scs.ryerson.ca}\thanks{Corresponding author}}
\author[1]{Alireza Sadeghian\thanks{asadeghi@ryerson.ca}}
\author[2]{Hamid Sadeghian\thanks{hsadeghian@mcvh-vcu.edu}}
\affil[1]{Dept. of Computer Science, Ryerson University, 350 Victoria Street, Toronto, ON M5B 2K3, Canada}
\affil[2]{Department of Neurology, Virginia Commonwealth University, 821 W Franklin St, Richmond, VA 23284, United States}

\providecommand{\keywords}[1]{\textbf{\textit{Keywords---}} #1}

\begin{document}

\maketitle

\begin{abstract}
In this paper, we analyze 48 signals of rest tremor velocity related to 12 distinct subjects affected by Parkinson's disease. The subjects belong to two different groups, high- and low-amplitude rest tremors, with four and eight subjects, respectively.
Each subject has been tested in four settings given by combining the use of deep brain stimulation and L-DOPA medication. We develop two main feature-based representations of such signals, which are obtained by considering (i) the long-term correlations and multifractal properties, and (ii) the power spectra.
Such feature-based representations are initially utilized for the purpose of characterizing the subjects under different settings.
Our results show that the effect of medication is clearly recognizable.
In addition, when medication is used, we observe a change in the related signals from anti-correlated to long-term, positively correlated.
In agreement with previous studies, we observe that deep brain stimulation does not significantly characterize neither of the two groups regardless of the adopted representation.
On the other hand, the medication effect yields statistically significant differences in both high- and low-amplitude tremor groups.
We successively test several different instances of the two feature-based representations of the signals in the setting of supervised classification and (nonlinear) feature transformation.
We consider three different classification problems, involving the recognition of (i) the presence of medication, (ii) the use of deep brain stimulation, and (iii) the membership to the high- and low-amplitude tremor groups.
Classification results show that the use of medication can be discriminated with higher accuracy, considering many of the feature-based representations.
Notably, we show that the best results are obtained with a parsimonious, two-dimensional representation encoding the long-term correlations and multifractal character of the signals.\\
\keywords{Long-term correlations; Multifractal spectra; Parkinsonian rest tremor; Classification; Feature transformation.}
\end{abstract}

\section{Introduction}

Long-memory processes describing complex systems \cite{beran2013long,yulmetyev2012correlations} and the analysis of long-term correlations (LTC) in related signals \cite{kechagias2015definitions} play an important role in many research contexts. For instance, we may consider investigations in EEG signals \cite{karkare2009investigating}, in the analysis subthalamic nucleus of patients with Parkinson’s disease \cite{hohlefeld2012long}, in DNA sequences \cite{papapetrou2014investigating}, in postural sway of humans \cite{delignieres2011transition}, and in construction engineering \cite{Pakrashi20131803}.
LTC are usually a primary source of fractality in signals -- in this paper, the terminology ``time series'' and ``signals'' are used interchangeably -- describing the observed variables of the system over time.
Fluctuations in a fractal time series present a power-law scaling, denoting thus the absence of a characteristic time/space scale describing the underlying system.
Multifractal time series are fractal time series that require more than one scaling exponent to be effectively described \cite{kantelhardt2009fractal}.
This happens when the underlying process is characterized by different scalings of the fluctuations, which in turn require a continuum of scaling exponents.
Several different methods have been developed in the past years to detect fractal and multifractal signatures. Among the many we can cite multifractal detrended fluctuation analysis (MF-DFA) \cite{kantelhardt2009fractal}, adaptive fractal analysis \cite{kirchner2014detrended}, wavelet transform modulo maxima \cite{PhysRevE.74.016103}, and wavelet leaders \cite{wavelet_leaders}.
Fractal and multifractal analysis of time series play a pivotal role in many scientific contexts, such as neuroscience and medicine in general \cite{west1994fractal,di2014fractals,di2015fractals}.
Just to mention a few, it is possible to cite applications in human gait analysis \cite{dutta2013multifractal}, background neuronal noise-like activity in human and mouse hippocampus \cite{serletis2012complexity}, analysis of cervical tissue samples \cite{10.1371/journal.pone.0108457}, MRIs for tumor characterization \cite{6548065}, EEG signals \cite{dbs2015}, protein contact networks \cite{mixbionets2}, and electromyograms for diagnosis of neuro-muscular diseases \cite{talebinejad2010multiplicative}.

Parkinson's disease (PD) is a neuro-degenerative disorder that targets the central nervous system.
PD is characterized by the progressive loss of dopaminergic neurons in the substantia nigra of the midbrain. The most evident symptoms associated with PD are tremors, bradykinesia, rigidity, and postural instability, while in more advanced stages of the disease other factors might be present such as different types of cognitive impairments (e.g., dementia) and changes in behavior and/or emotional states \cite{beuter2001effect,Haeri2005311}. The causes of PD are, however, still largely unknown.
This has led to multi-disciplinary research involving, for instance, the use of artificial neural networks for the purpose of prediction of related signals \cite{wu2010prediction,engin2007classification} and mutual information based methods for detecting upper limb motor dysfunction \cite{de2011use}.
Deep brain stimulation (DBS) \cite{mcintyre2010network,mayberg2005deep,slow2014deep} is neurosurgical procedure that involves a surgical intervention to implant electrodes in brain areas suitable for receiving electrical impulses. DBS proved to be effective in the treatment of PD and other diseases, such as obsessive-compulsive disorders \cite{bronstein2011deep}.

In this paper, we study 48 signals recorded from 12 subjects affected by PD \cite{beuter2001effect}. Tremor signals are recorder by means of a velocity laser targeted to their index finger. For each subject, signal recording is performed in four different settings, given by the combination of the use of DBS and L-DOPA medication.
The original experiment \cite{beuter2001effect,titcombe2001dynamics} was performed on a larger set of 16 subjects. However, not all subjects were recorded in the four aforementioned conditions. Therefore, here we consider four subjects affected by high-amplitude tremors and eight affected by low-amplitude tremors.
Using the same data, Yulmetyev et al. \cite{yulmetyev2006regular} performed a comprehensive analysis by using the statistical theory of non-Markov processes and flicker-noise spectroscopy.
In addition, the attenuation effects of DBS on locomotion and tremor over different time scales were further investigated by Beuter and Modolo \cite{beuter2009}, which developed a computational model of biological neural networks.

Here we proceed on a different route by using LTC and multifractal spectra (MFS) as representation and analysis tools.
To our knowledge, a characterization of such signals for the purpose of discrimination in terms of LTC and MFS has never been developed in the literature. LTC and multifractal properties are derived here by means of the MF-DFA technique.
We develop two main low-dimensional, feature-based representations (FBRs) of such signals, which are obtained by considering (i) the LTC and multifractal properties and (ii) the power spectra. The power spectra are principally used for comparison purposes, since they have been already analyzed in detail by Beuter et al. \cite{beuter2001effect}.
The FBRs are initially utilized for the purpose of characterizing the subjects under different test settings.
All signals present a clear multifractal signature and different forms of LTC.
Our results show that the effect of medication is clearly recognizable in the signals. In addition, the use of medication indicates a qualitative change of LTC from anti-persistent to persistent.
In agreement with previous studies \cite{beuter2001effect}, we show that DBS does not characterize neither of the two groups, regardless of the adopted FBR. On the other hand, the medication effect yields statistically significant differences in both high- and low-amplitude tremor groups.
We successively test several different instances of the two FBRs of the signals in the setting of supervised classification and (nonlinear) feature transformation.
We consider three different classification problems involving the recognition of (i) the presence of medication, (ii) the use of deep brain stimulation, and (iii) the high- and low-amplitude tremor groups.
Classification results show that the use of medication can be discriminated with higher accuracy, considering many of the FBRs.
In particular, we show that the best results are obtained with the herein developed parsimonious, two-dimensional representation encoding the LTC and multifractal character of the signals.
Overall, our results highlight the usefulness of LTC and multifractal signatures in the analysis Parkinsonian rest tremors.

The remainder of this paper is structured as follows.
In Section \ref{sec:fbrepr}, we introduce the 48 signals, describing the developed FBRs. For details related to the experimental setting on which the original signals were obtained the readers are referred to the original study \cite{beuter2001effect}.
In Section \ref{sec:results}, we present and discuss the results related to the developed FBRs of such signals.
In Section \ref{sec:conclusions}, we draw the conclusions and discuss future directions.
The paper includes Appendix \ref{sec:mfdfa} that provides the essential details about MF-DFA.

\section{Feature-based representation of the rest tremor signals}
\label{sec:fbrepr}

The original experiment \cite{beuter2001effect} consisted of recording rest tremor velocity from 16 subjects affected by Parkinson's disease.
Rest tremor was recorded via a velocity laser under four different main conditions, given by combining the use of medication (L-DOPA) and high-frequency DBS.
Participants received DBS of the internal globus pallidus, the subthalamic nucleus, or the ventrointermediate nucleus of the thalamus.
Unfortunately, not all subjects were recorded in all four conditions and only for 12 subjects all four recording conditions (i.e., medication Off--On, DBS Off--On) are available. Among those 12 subjects, four belong to the high-amplitude rest tremor group (originally denoted as g2, s6, s7, and s8) while the remaining eight to the low-amplitude rest tremor group (originally denoted as g9, g10, g11, g12, g13, s14, s15, and s16).
In the following, we study the related 48 signals of rest tremors.

\begin{figure}[ht!]
\centering

\subfigure[H-MFSW representation.]{
\includegraphics[viewport=0 0 343 242,scale=0.6,keepaspectratio=true]{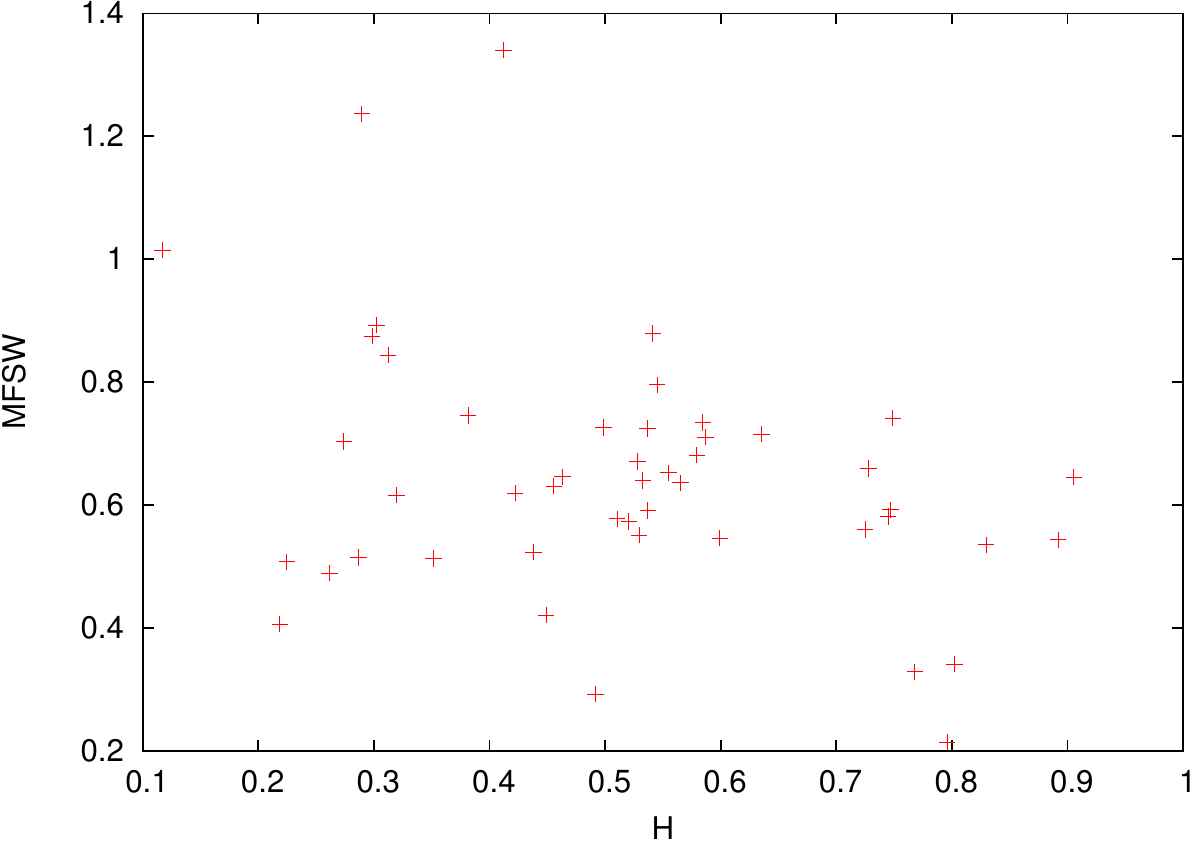}
\label{fig:HW}}
~
\subfigure[PCA of MFS coefficients.]{
\includegraphics[viewport=0 0 345 242,scale=0.6,keepaspectratio=true]{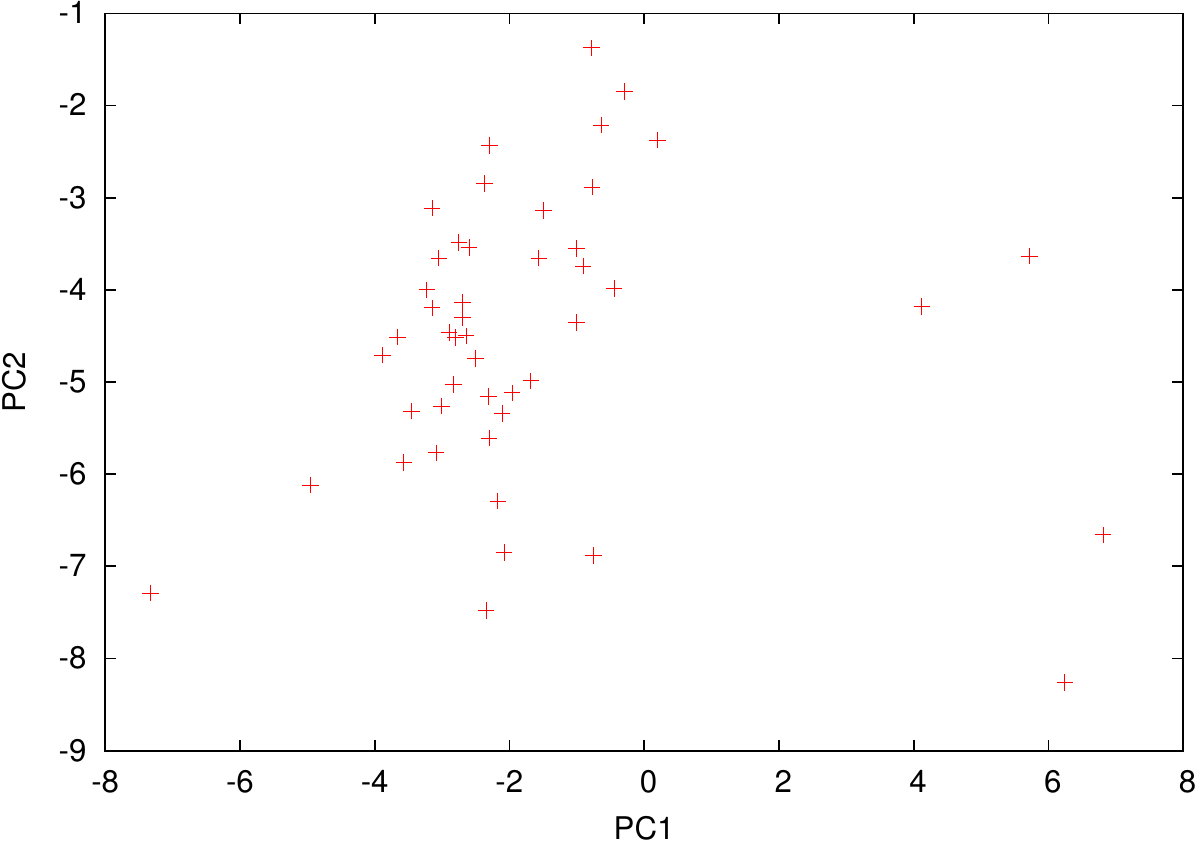}
\label{fig:MFS_PCA}}

\subfigure[Kernel PCA of MFS coefficients.]{
\includegraphics[viewport=0 0 348 242,scale=0.6,keepaspectratio=true]{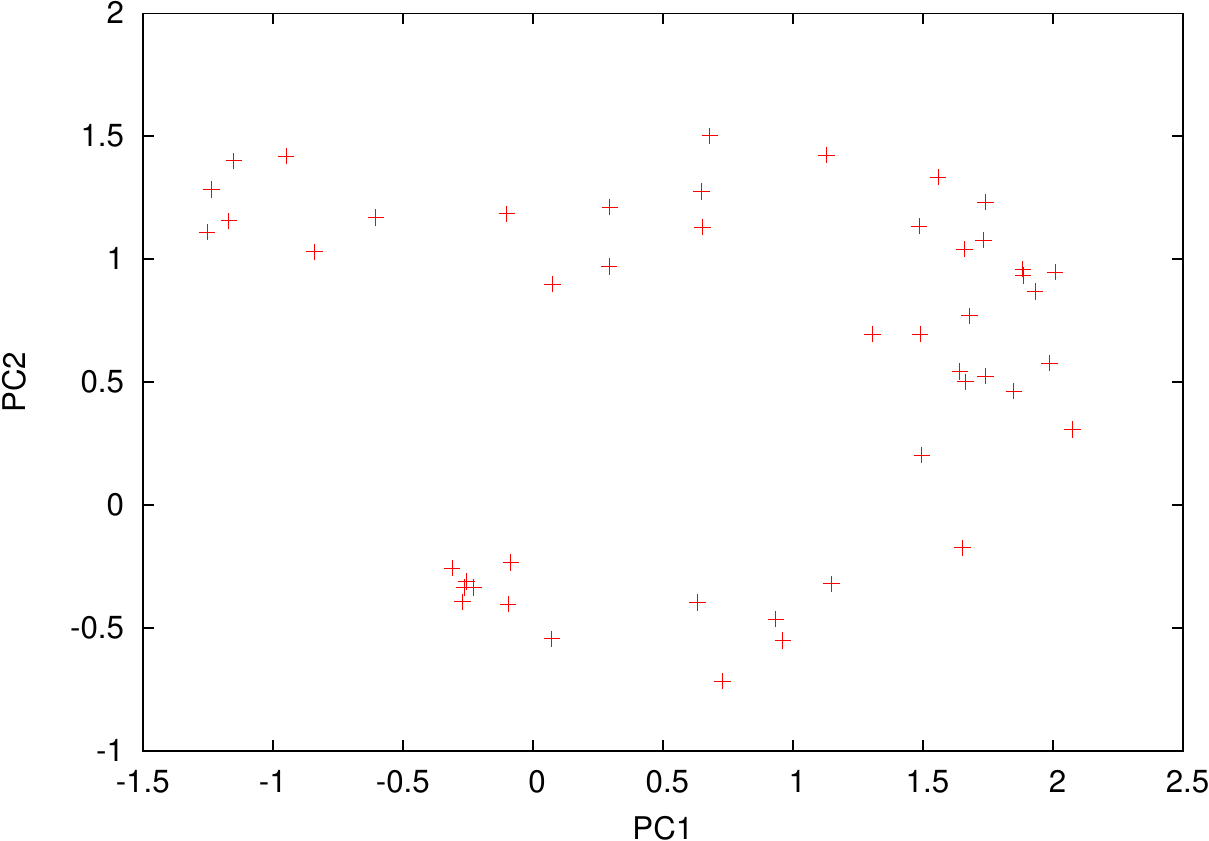}
\label{fig:MFS_kPCA}}

\caption{Representations using Hurst exponent and (nonlinear) transformation of MFS coefficients.}
\label{fig:MFS_REPR}
\end{figure}

These 48 signals are represented here according to two main FBRs, i.e., as numeric vectors.
Notably, we represent such signals by using (i) LTC and MFS properties and (ii) the estimated power spectra.
In the first case, we develop three, low-dimensional representations by means of the coefficients derived via MF-DFA -- see Appendix \ref{sec:mfdfa} for technical details.
MF-DFA is executed by considering 40 equally-spaced (temporal) scales in-between 16 and 512; we remove quadratic (local) trends; we consider 101 equally-spaced values in-between -5 and 5 for the fluctuation weighting factor $q$. Such settings are rather standard and proved to be effective in our study.
The FBR shown in Fig. \ref{fig:HW} uses only the Hurst exponent (H) (\ref{eq:Fq}) and the multifractal spectrum width (MFSW) (\ref{eq:width}) elaborated from the available time series. In the following we denote such a representation as H-MFSW.
Next, we consider the information provided by the entire MFS. Specifically, we initially represent each time series as a high-dimensional vector encoding the domain (\ref{eq:mfs_domain}) and co-domain of the calculated MFS (\ref{eq:mutifractal_spectrum}).
Such high-dimensional vectors are then transformed by considering both principal component analysis (PCA) and its nonlinear extension \cite{hoffmann2007kernel}, known as kernel PCA (kPCA); where a Gaussian kernel is adopted.
We noted that the first four principal components (PCs) usually explain more than 80\% of the data variance, so in both cases they are retained for the embedding.
The MFS coefficients transformed via the linear PCA are denoted as MFS-PCA, while MFS-kPCA is used to denote the four-dimensional vectors obtained by the corresponding nonlinear transformation -- see respectively Fig. \ref{fig:MFS_PCA} and Fig. \ref{fig:MFS_kPCA}.

\begin{figure}[ht!]
\centering

\subfigure[PCA of power spectra.]{
\includegraphics[viewport=0 0 348 242,scale=0.6,keepaspectratio=true]{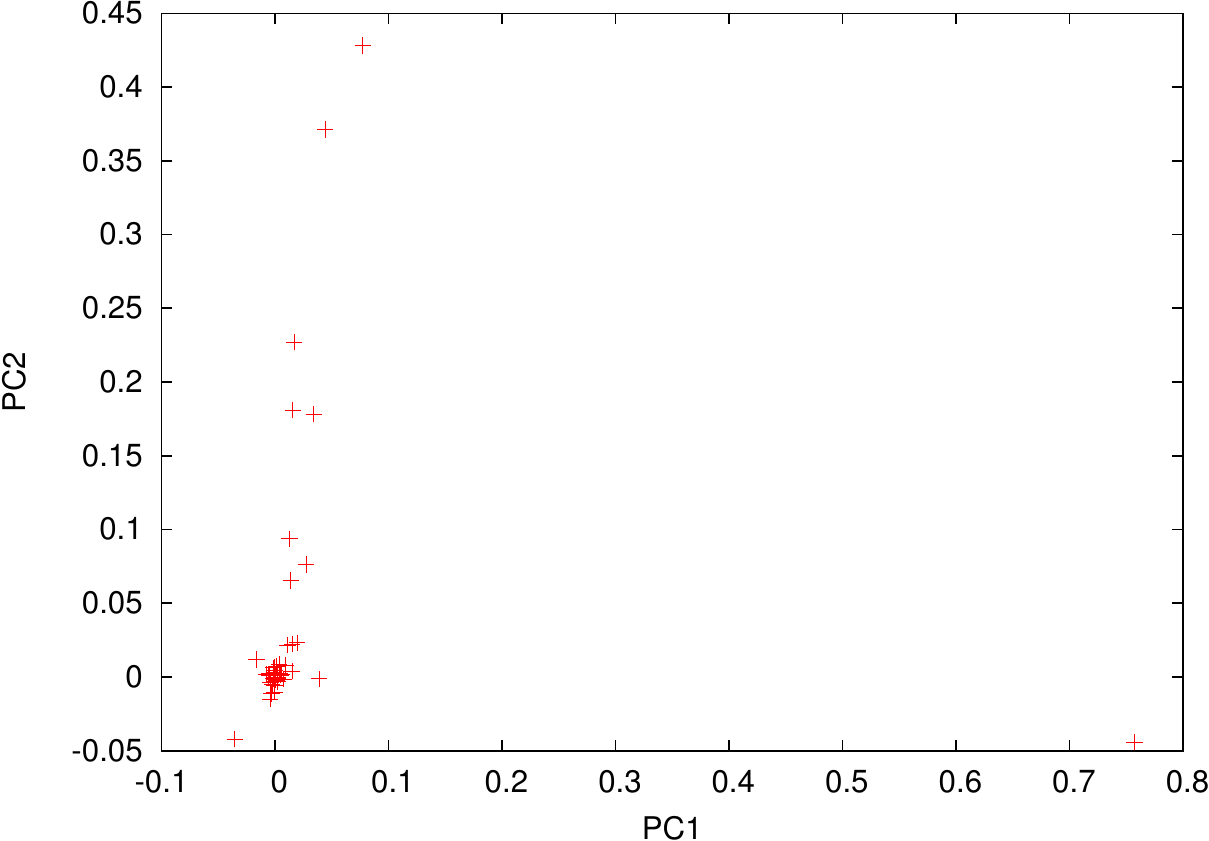}
\label{fig:power_PCA}}
~
\subfigure[Kernel PCA of power spectra.]{
\includegraphics[viewport=0 0 348 242,scale=0.6,keepaspectratio=true]{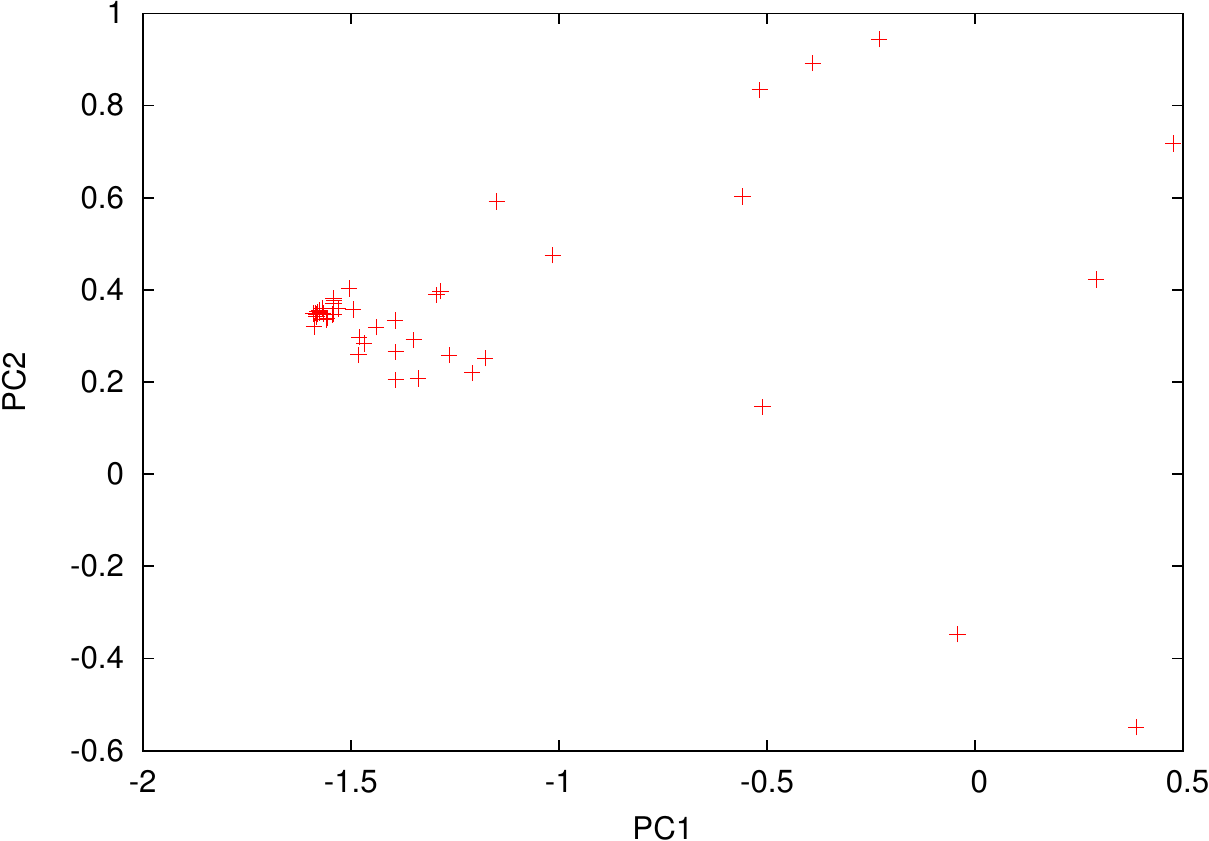}
\label{fig:power_kPCA}}

\caption{Representations using (nonlinear) transformation of the power spectra coefficients.}
\label{fig:power_REPR}
\end{figure}

Power spectrum is estimated using the well-known Welch's method \cite{1161901}.
We initially represented each time series as a high-dimensional vector containing the amplitude values at 1025 (normalized) frequencies.
Such high-dimensional representations are then transformed by using PCA and kPCA; four dimensions are retained as in the previous case.
Fig. \ref{fig:power_REPR} shows the first two PCs of such representations, denoted in following as POWER-PCA and POWER-kPCA, respectively.

Please note that the herein developed FBRs of the original signals are not ``labeled'' yet, in the sense that here we did not consider any specific discrimination problem.
In Sec. \ref{sec:results} we will deal with several characterizations and discrimination problems aptly conceived over such FBRs.


\section{Analysis of experimental results}
\label{sec:results}

Our main objective here is to show that LTC and MFS can be adopted as effective features when representing the considered biomedical signals.
We contrast the results with FBRs obtained from the more conventional power spectrum. We show that the two FBRs provide comparable results, although the former allows also to infer important qualitative information regarding the signals.
We proceed by analyzing the different FBRs developed in Sec. \ref{sec:fbrepr} by first taking into account the independent contribution of the features, while considering the contribution of each feature in the discrimination of the two groups of high- and low-amplitude rest tremor subjects (Sec. \ref{sec:char}). Successively, we evaluate the FBRs in terms of recognition performance over three suitably-defined classification problems (Sec. \ref{sec:class}).

\subsection{Characterization and statistical analysis}
\label{sec:char}

Let us first take into account the results delivered in Table \ref{tab:diff}. For the first FBR, we assessed the Hurst exponent (H) and MFSW, while for the second one we considered the first two PCs obtained from the PCA and kPCA, respectively. We categorize the data by considering the use of medication, the use of DBS, and subjects with high- and low-amplitude tremors.
$p$-values are obtained by evaluating the t-test over each categorization, where we consider the usual 5\% as the threshold.
The use of medication can be suitably recognized when taking into account most of the considered variables.
In fact, both H and MFSW result in a statistically significant discriminator for such a categorization.
The use of DBS, instead, does not allow for any statistically significant discrimination -- in terms of t-test.
This suggests that, according to the FBRs that we used, DBS does not have a significant global impact on the subjects of the two groups. This aspect calls for further analysis, since in fact DBS is typically used in many clinical settings.
Finally, only two of the six considered variables (namely, MFSW and POWER-kPC1) produce statistically significant differences when considering the characterization in terms of high/low amplitude rest tremors.

\begin{figure*}[htp!]
 \centering
 \includegraphics[viewport=0 0 360 241,scale=1.2,keepaspectratio=true]{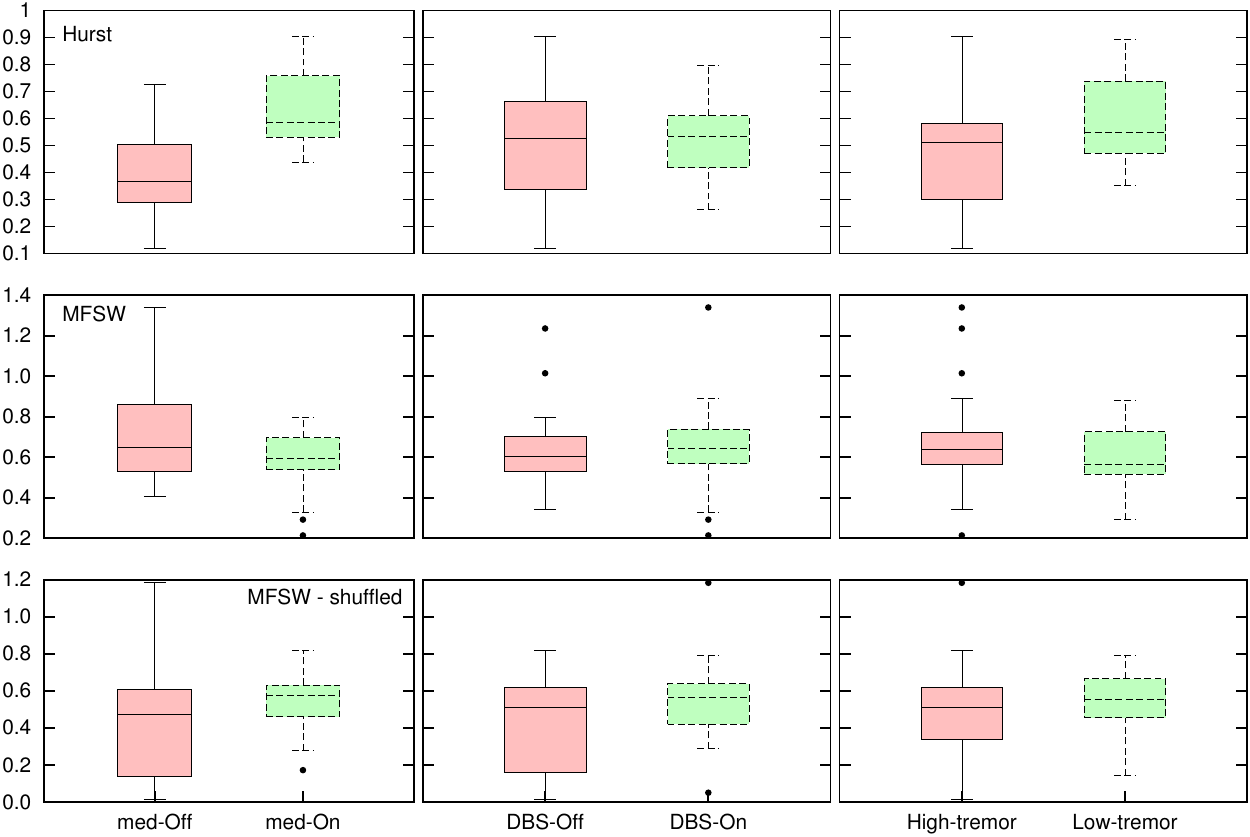}
 \caption{Box-plots representing the statistics of Hurst and MFSW when considering the use of medication, DBS, and the amplitude of the rest tremors.}
 \label{fig:hurst_mfsw}
\end{figure*}

Fig. \ref{fig:hurst_mfsw} offers a visual representation of such statistics (for H and MFSW only) for the three different categories.
According to the $p$-values in Table \ref{tab:diff}, LTC properties of the signals yields statistically significant differences only when considering medications.
It can be noted that the use of medication changes the signal LTC properties toward positively correlated, while in absence of medication the signals are clearly anti-correlated (upper panels). In our opinion, this is an interesting aspect that indicates the need for further developments in future research studies.
All signals are multifractal, with a relevant multifractal signature quantified by the MFSW (middle panels).
It is worth noting that the multifractal signature is sufficiently preserved after shuffling the time series (lower panels), suggesting that LTC are not the only source of the observed multifractality. In fact, shuffling destroys LTC and any deterministic trend that might influence the degree of multifractality of the series.
In conclusion, we note that differences between the MFSW of the original and shuffled time series are statistically significant ($p<0.0008$). Nonetheless, a more detailed verification shows that when medication is Off (On) differences are (not) statistically significant between the original and shuffled time series, $p<0.0001$ ($p<0.5251$); a similar scenario holds for the use of DBS, $p<0.0022$ ($p<0.1004$), and for high- (low-) amplitude tremors, $p<0.0037$ ($p<0.0816$). 
\begin{table*}[thp!]\footnotesize
\caption{$p$-values -- statistically significant results are reported in bold. The columns named ``med-Off / med-On'' and ``DBS-Off / DBS-On'' consider differences between all subjects in the respective settings, while the column ``High-tremor / Low-tremor'' the differences between the two groups taking into account all combinations of DBS and medication.}
\begin{center}
\begin{tabular}{|c|c|c|c|}
\hline
\textbf{Feature} & \textbf{med-Off / med-On} & \textbf{DBS-Off / DBS-On} & \textbf{High-tremor / Low-tremor} \\
\hline
H & $\bm{p<0.0001}$ & $p<0.9291$ & $p<0.3860$ \\ 
MFSW & $\bm{p<0.0272}$ & $p<0.7727$ & $\bm{p<0.0134}$ \\ 
\hline
POWER-PC1 & $p<0.2427$ & $p<0.4869$ & $p<0.1122$ \\ 
POWER-PC2 & $\bm{p<0.0114}$ & $p<0.4879$ & $p<0.0507$ \\ 
POWER-kPC1 & $\bm{p<0.0001}$ & $p<0.4721$ & $\bm{p<0.0059}$ \\ 
POWER-kPC2 & $p<0.4614$ & $p<0.0675$ & $p<0.6932$ \\ 
\hline
\end{tabular}
\label{tab:diff}
\end{center}
\end{table*}

Let us now take into account the results in Table \ref{tab:diff2}, which show the contribution of DBS and medication on the characterization of subjects having high- and low-amplitude tremors, respectively.
As suggested by the original experiments conducted by Beuter et al. \cite{beuter2001effect}, DBS seems not to have a statistically significant impact on such a characterization.
In fact, in both groups, regardless of the considered feature, we obtain $p$-values far from denoting statistically significant results.
On the other hand, when considering the impact of medication on the two groups, we obtain some statistically significant results.
For the group having high-amplitude rest tremors (indicated as ``High-tremor'' in the table), features H, POWER-PC2, and POWER-kPC1 produce statistically significant results, while for the low-amplitude group we have H and POWER-kPC1.

\begin{table}[thp!]\footnotesize
\caption{$p$-values are calculated to assess differences of each group (high and low tremor subjects) when considering the effects of DBS and medication. DBS does not produce statistically significant differences, while medication typically does yield statistically significant differences (reported in bold).}
\begin{center}
\begin{tabular}{|c|c|c|c|}
\hline
\textbf{Feature} & \textbf{High-tremor} & \textbf{Low-tremor} \\
\hline
\multicolumn{3}{|c|}{\textbf{DBS-Off / DBS-On}} \\
\hline
H & $p<0.9587$ & $p<0.8444$ \\ 
MFSW & $p<0.6068$ & $p<0.4481$ \\ 
\hline
POWER-PC1 & $p<0.4159$ & $p<0.0692$ \\ 
POWER-PC2 & $p<0.6833$ & $p<0.4167$ \\ 
POWER-kPC1 & $p<0.3856$ & $p<0.9868$ \\ 
POWER-kPC2 & $p<0.1485$ & $p<0.2881$ \\ 
\hline
\multicolumn{3}{|c|}{\textbf{med-Off / med-On}} \\
\hline
H & $\bm{p<0.0001}$ & $\bm{p<0.0001}$ \\ 
MFSW & $p<0.0596$ & $p<0.1753$ \\ 
\hline
POWER-PC1 & $p<0.2402$ & $p<0.8188$ \\ 
POWER-PC2 & $\bm{p<0.0382}$ & $p<0.0823$ \\ 
POWER-kPC1 & $\bm{p<0.0001}$ & $\bm{p<0.0139}$ \\ 
POWER-kPC2 & $p<0.8006$ & $p<0.2780$ \\ 
\hline
\end{tabular}
\label{tab:diff2}
\end{center}
\end{table}

\subsection{Classification of rest tremor signals}
\label{sec:class}

In order to evaluate the discrimination capability of each FBR, we have considered three supervised classification problems.
We face the problem of discriminating between the two groups (high- and low-amplitude tremors) and recognizing the use of medication and DBS, respectively.
We have chosen the well-known support vector machine (SVM) as supervised classification systems, configured with a Gaussian kernel. Notably, we used a version known as C-SVM, where C is a hyper-parameter controlling the complexity of the resulting model -- in SVM the structural complexity of the model is measured by considering the number of support vectors (SVs) computed during the training stage \cite{theodoridis2006pattern}.
Both hyper-parameters, i.e., C and the width of the Gaussian kernel, are determined by preliminary tests using a typical grid search scheme.
Since our dataset, regardless of the adopted FBR, is limited to 48 patterns, we tested the recognition capability of C-SVM according to the leave-one-out setting: each pattern is tested by learning a C-SVM model on the remaining 47 patterns.
Table \ref{tab:classification} summarizes the results for all three classification problems.
We report for each classification problem the results obtained with the five FBRs.
For POWER-PCA and POWER-kPCA we use, in both cases, the first 2, 3, and 4 PCs in order to evaluate the performances by varying the dimensionality of the representation.
We show the number of errors (for each class), the area under the receiver operating characteristic curve (AUC), and the average number of SVs as an indicator of C-SVM model structural complexity.
Generally, results show that effect of medication allows more accurate classification.
We note that the best result (AUC is 0.85) is obtained with the parsimonious, two-dimensional representation denoted as H-MFSW; the per-class errors are also more balanced with respect to the results obtained with the two-dimensional version of POWER-kPCA.
In general, results for the last two problems, namely ``DBS-Off / DBS-On'' and ``High-tremor / Low-tremor'', are not convincing -- we obtain results compatible with a random classifier.
This fact confirms that the effect of medication on the 48 subjects seems to be more characterizing, allowing for a good discrimination regardless of the use of DBS or the membership to the high- or low-amplitude rest tremor groups.
\begin{table}[th!]\footnotesize
\caption{Classification results with leave-one-out on the FBRs of the input signals. Three different classification problems are faced by considering several feature-based, low-dimensional representations: recognition of (i) medication Off--On, (ii) DBS Off--On, and (iii) high--low amplitude tremor.}
\begin{center}
\begin{tabular}{|c|c|c|c|c|}
\hline
\textbf{Representation} & \textbf{Dimension} & \textbf{Errors} & \textbf{AUC} & \textbf{SVs} \\
\hline
\multicolumn{5}{|c|}{\textbf{\textit{med-Off / med-On}}} \\
\hline
H-MFSW & 2 & 7 (5/24, 2/24) & \textbf{0.85} & 35.2 \\ 
MFS-PCA & 4 & 16 (8/24, 8/24) & 0.67 & 44.2 \\ 
MFS-kPCA & 4 & 16 (10/24, 6/24) & 0.67 & 31.6 \\ 
\hline
POWER-PCA & 2 & 22 (22/24, 0/24) & 0.54 & 46.0 \\
POWER-PCA & 3 & 22 (22/24, 0/24) & 0.54 & 46.5 \\
POWER-PCA & 4 & 23 (23/24, 0/24) & 0.52 & 46.5 \\
POWER-kPCA & 2 & 13 (13/24, 0/24) & 0.73 & 31.5 \\
POWER-kPCA & 3 & 9 (9/24, 0/24) & 0.81 & 31.1 \\
POWER-kPCA & 4 & 8 (8/24, 0/24) & 0.83 & 31.1 \\
\hline
\multicolumn{5}{|c|}{\textbf{\textit{DBS-Off / DBS-On}}} \\
\hline
H-MFSW & 2 & 48 (24/24, 24/24) & 0.00 & 47.0 \\ 
MFS-PCA & 4 & 29 (14/24, 15/24) & 0.40 & 47.0 \\ 
MFS-kPCA & 4 & 26 (8/24, 18/24) & 0.46 & 44.6 \\ 
\hline
POWER-PCA & 2 & 48 (24/24, 24/24) & 0.00 & 46.9 \\
POWER-PCA & 3 & 48 (24/24, 24/24) & 0.00 & 46.9 \\
POWER-PCA & 4 & 48 (24/24, 24/24) & 0.00 & 46.9 \\
POWER-kPCA & 2 & 23 (3/24, 21/24) & \textbf{0.52} & 45.2 \\
POWER-kPCA & 3 & 23 (3/24, 21/24) & 0.52 & 45.3 \\
POWER-kPCA & 4 & 23 (3/24, 21/24) & 0.52 & 45.9 \\
\hline
\multicolumn{5}{|c|}{\textbf{\textit{High-tremor / Low-tremor}}} \\
\hline
H-MFSW & 2 & 16 (16/16, 0/32) & 0.50 & 32.7 \\ 
MFS-PCA & 4 & 21 (14/16, 7/32) & 0.45 & 39.2 \\ 
MFS-kPCA & 4 & 15 (12/16, 3/32) & 0.58 & 32.2 \\ 
\hline
POWER-PCA & 2 & 16 (16/16, 0/32) & 0.50 & 31.3 \\
POWER-PCA & 3 & 16 (16/16, 0/32) & 0.50 & 31.5 \\
POWER-PCA & 4 & 16 (16/16, 0/32) & 0.50 & 32.2 \\
POWER-kPCA & 2 & 13 (12/16, 1/32) & \textbf{0.61} & 29.2 \\
POWER-kPCA & 3 & 14 (13/16, 1/32) & 0.58 & 30.2 \\
POWER-kPCA & 4 & 14 (13/16, 1/32) & 0.58 & 28.9 \\
\hline
\end{tabular}
\label{tab:classification}
\end{center}
\end{table}

\section{Conclusion and future directions}
\label{sec:conclusions}

We have studied 12 subjects divided in two groups affected by high- and low-amplitude Parkinsonian rest tremors, respectively.
Each subject has been tested in four settings, given by combining the use of deep brain stimulation and L-DOPA medication for relieving tremors and other symptoms.
As a result, our initial dataset was formed by 48 signals related to the rest tremors measured via a velocity laser pointing at the index finger of the participants.
We developed two main feature-based representations of such signals, which have been obtained by considering (i) the long-term correlations and multifractal properties and (ii) the power spectra.
Initially, we have used such feature-based representations for the purpose of characterizing the subjects under different test settings.
We have shown that the effect of medication is clearly recognizable in the represented signals.
In agreement with previous studies, we have found that deep brain stimulation does not discriminate the two groups, regardless of the adopted representation. On the other hand, our results suggested that the effects of medication produce statistically significant differences in both groups having high and low-amplitude tremor.
We successively tested several different instances of the two feature-based representations of the signals in the setting of supervised classification and (nonlinear) feature transformation.
Three different classification problems have been considered, involving the recognition of (i) the presence of medication, (ii) the use of deep brain stimulation, and (iii) the membership to the high- and low-amplitude tremor groups.
Classification results demonstrated that the use of medication can be discriminated with higher accuracy.
Interestingly, the best results were obtained with a parsimonious, two-dimensional representation encoding the long-term correlations and multifractal character of the original signals.
We believe that such results could be potentially useful to neuroscientists, suggesting the potential of using LTC and multifractal signatures for the analysis Parkinsonian rest tremors.

Both long-term correlations and multifractal properties have been derived by using the ``classical'' multifractal detrended fluctuation analysis.
Future research works might exploit the direct estimation of the multifractal spectrum \cite{PhysRevLett.62.1327,chhabra1989direct} and related time-dependent Hurst exponent for the same aim of signal characterization and discrimination.
Direct estimation of the multifractal spectrum might provide a compelling alternative for this purpose, especially when processing time series that span a limited time frame.

\appendix
\section{Multifractal detrended fluctuation analysis}
\label{sec:mfdfa}

Full details about the MF-DFA procedure are provided in Ref. \cite{kantelhardt2002multifractal}.
Given a time series $x_k$ of length $N$ with compact support, the following steps are performed.
Let
\begin{equation}
Y(i) = \sum_{k=1}^i \left[ x_k - \langle x \rangle \right], \;\; i = 1, \dotsc, N,
\end{equation}
be the profile, which is successively divided in $N_s = \text{int}(N/s)$ non-overlapping segments of equal length $s$. 
Since $N$ might not be a multiple of $s$, the operation is repeated by starting from the opposite end, obtaining thus a total of $2N_s$ segments.

Successively, a local detrending operation is executed by calculating a polynomial fit on each of the $2N_s$ segments.
Then the local variance is determined as
\begin{equation}
F^2(\nu,s) = \frac{1}{s} \sum_{i=1}^s \bigg\{ Y[(\nu-1)s+i] - y_\nu(i) \bigg\}^2,
\end{equation}
for each segment $\nu = 1,\dotsc,N_s$ and
\begin{equation}
F^2(\nu,s) = \sum_{i=1}^s \bigg\{ Y[N-(\nu - N_s)s+i] - y_\nu(i)\bigg\}^2
\end{equation}
for $\nu = N_s +1,\dotsc, 2N_s$, where $y_\nu(i)$ is the fitted polynomial in segment $\nu$.
The order $m$ of the fitting polynomial, $y_\nu(i)$, affects the capability of removing trends in the series; it has to be tuned according to the expected trending order of the time series.
The $q$th-order average of the variance over all segments is evaluated as
\begin{equation}
\label{eq:Fq}
F_q(s) = \bigg\{ \frac{1}{2N_s} \sum_{\nu=1}^{2N_s} \left[ F^2(\nu,s)\right]^{q/2} \bigg\}^{1/q},
\end{equation}
with $q \in \mathbb{R}$. The $q$-dependence of the fluctuations function (\ref{eq:Fq}) allows to highlight the contributions of both high and low fluctuation magnitudes. Notably, for $q > 0$ only the larger fluctuations have higher impact in Eq.~\ref{eq:Fq}; conversely, for $q < 0$ the impact of the smaller fluctuations is enhanced. The case $q = 0$ cannot be computed by means of the average in Eq.~\ref{eq:Fq} and so a logarithmic form has to be used.
The last steps are repeated for different scale sizes, $s$.

The scaling behavior of the fluctuations can be determined by analyzing the slope of the doubly-logarithmic plot of $F_q(s)$ versus $s$, computed for each value of $q$. If the series $x_i$ is long-term correlated, then $F_q(s)$ is approximated -- for large values of $s$ -- by the power-law form:
\begin{equation}
\label{eq:Fqshq}
F_q(s) \sim s^{H(q)}.
\end{equation}

The $H(q)$ exponent is the generalization of the Hurst exponent, $H$, which is obtained for $q=2$. When $H>1/2$ the time series possesses long-term, positive correlations; for $H<1/2$ the series is anti-correlated; $H=1/2$ indicates that the series is compatible with uncorrelated noise -- e.g., white Gaussian noise. 
An equivalent scaling over all fluctuation magnitudes indicates that $H(q)$ is independent from $q$, suggesting that the series is monofractal. On the contrary, when the small fluctuations scale differently from the large ones, then the series can be considered as multifractal.

Starting from Eq.~\ref{eq:Fq} and using Eq.~\ref{eq:Fqshq}, it is possible to define
\begin{equation}
\label{mass_exponent}
Z_q(s)=\sum_{\nu=1}^{N/s} [ F(\nu,s)]^q \sim s^{qH(q) - 1},
\end{equation}
where
\begin{equation}
\label{eq:tauq}
\tau(q) = qH(q) - 1
\end{equation}
is the $q$-order mass exponent (also called R\'{e}nyi scaling exponent) of the generalized partition function, $Z_q(s)$.
The multifractal spectrum, denoted as $f(\cdot)$, provides a compact description of the multifractal character of the time series.
Such a function cab be obtained via the Legendre transform of $\tau(q)$,
\begin{equation}
\label{eq:mutifractal_spectrum}
f(\alpha) = q\alpha - \tau(q),
\end{equation}
where $\alpha$ is equal to the derivative $\tau'(q)$ -- it corresponds to the H\"older exponent, also called singularity exponent.
Using Eq.~\ref{eq:tauq}, it is possible to express the generalized Hurst exponent, $H(q)$, in terms of $\alpha$ and $f(\alpha)$,
\begin{equation}
\label{eq:mfs_domain}
\alpha = H(q) + qH'(q) \;\; \text{and} \;\; f(\alpha) = q[\alpha - H(q)] + 1.
\end{equation}

The multifractal spectrum (\ref{eq:mfs_domain}) encodes important information regarding the degree of multifractality and the specific sensitivity of the time series to fluctuations of high/low magnitudes.
Let $q_{-}$ and $q_{+}$ be, respectively, the lower and upper values chosen for the $q$ range. The width of the support of $f(\cdot)$, defined as
\begin{equation}
\label{eq:width}
\Delta\alpha=\alpha(q_{-})-\alpha(q_{+}),
\end{equation}
offers an important quantitative indicator of the multifractal signature that is present in the data.

\bibliographystyle{abbrvnat}
\bibliography{/home/lorenzo/University/Research/Publications/Bibliography.bib}
\end{document}